%% file: CLANEK.TEX
\documentstyle{ioplppt}
\begin{document}
\eqnobysec
\input makra1
\jl{6}
\title{Curvature invariants in type {\it{N}} spacetimes}
\author{J. Bi\v c\' ak{\ftnote{3}{E-mail: bicak@mbox.troja.mff.cuni.cz}}
  and V. Pravda{\ftnote{4}{E-mail: pravda@otokar.troja.mff.cuni.cz}}}
\address{
Department of Theoretical Physics, Faculty of Mathematics  and Physics,\\
Charles University, V Hole\v sovi\v ck\' ach 2,  180 00 Prague 8, Czech
Republic}
\begin{abstract}
Scalar curvature invariants are studied in type {\it N} solutions of vacuum
Einstein's equations with in general non-vanishing cosmological constant
$\Lambda$. Zero-order invariants which include only the metric and
Weyl (Riemann) tensor either vanish, or are constants depending on
$\Lambda$. Even all higher-order invariants containing covariant derivatives
of the Weyl (Riemann) tensor are shown to be trivial if a type {\it N} spacetime admits a
non-expanding and non-twisting null geodesic congruence.

However, in the case of expanding type {\it N} spacetimes
we discover a non-vanishing scalar invariant which is quartic in the
second derivatives of the Riemann tensor.

We use this invariant to demonstrate that both linearized and the third
order type {\it N} twisting solutions recently discussed in literature contain
singularities at large distances and thus cannot describe radiation
fields outside bounded sources.
\end{abstract}
\pacs{0420, 0430}

\section{Introduction}

The Petrov algebraic classification of the Weyl tensor and the asymptotic
forms of radiative fields of spatially bounded sources
(peeling theorem) demonstrate that solutions of type {\it N} play a fundamental
role in the theory of gravitational radiation. (For a recent review of exact
approaches to radiative spacetimes, see e.g. \cite{Bicak} and references therein.)

All solutions of the vacuum Einstein equations with in general non-zero
cosmological constant $\Lambda$, which are of type {\it N} with a  non-twisting
null congruence, are known \cite{Kramer}, \cite{GP}, \cite{Pod}. In the twisting case, only one (Hauser's)
solution is available \cite{Kramer}. Although far-zone radiation fields of bounded sources
are approximately of type {\it N}, no known exact type {\it N} solution is asymptotically
flat.

Recently, an interesting discussion  appeared in literature \cite{Step1}, \cite{Step2}, \cite{Finley} in which
twisting type {\it N} vacuum field equations were solved approximately. Stephani \cite{Step2}
argued, within a linear theory, that no solutions regular outside a bounded
source, exist. Finley, Pleba\` nski and Przanowski \cite{Finley}  iterated the solution
up to the third order and concluded that their iterative procedure leads to
the regular solutions, so that "it seems that the twisting, type {\it N} fields
{\it{can}} describe a radiation field outside a bounded sources."

In neither of these works, however, the arguments are really compelling since
 singularities are not analyzed properly. (For general treatments
of singularities, see e.g. {\cite{Clarke}}, {\cite{Ellis}}.)  The authors study only
the behaviour of invariants with respect to gauge transformations which leave
invariant coordinate system and field equations for type {\it N} metrics because
for type {\it N} vacuum spacetimes with $\Lambda$ all scalar invariants of the Weyl or Riemann
tensor are trivial - they either vanish or are constants depending on $\Lambda$.

It is known that for a vacuum pp-wave, which is a very special case of
 a type {\it N} metric, in fact all curvature invariants of {\it {any}}
order, i.e., invariants depending also on  covariant {\it derivatives} of
the Weyl (or Riemann) tensor, vanish \cite{Jord}, \cite{Schmidt}. However, nothing
appears to be known about higher-order invariants in more general
type {\it N} spacetimes. The problem of such higher-order invariants is addressed in the present
paper\footnote{It is partially based on the thesis {\cite{Pravda}}}.

We define an invariant of the metric $g_{\na \nb} (x^\ng) $  of the order $ k$
as a non-constant scalar function $I(g_{\na \nb} ,g_{\na \nb ,\ng_1},\dots ,
g_{\na \nb ,\ng_1 \dots \ng_{k+2}}  )$
which satisfies
\BE
I(g_{\na \nb} ,g_{\na \nb ,\ng_1},\dots ,
g_{\na \nb ,\ng_1 \dots \ng_{k+2}}  ) = I({g'}_{\na \nb} ,{g'}_{\na \nb ,\ng_1},\dots ,
{g'}_{\na \nb ,\ng_1 \dots \ng_{k+2}}  )
\EE
under a spacetime diffeomorphism $x^\na \rightarrow  {x'}^\na = {x'}^\na (x^\nb) $.
It can be proved (see e.g. \cite{Klinger} ) that any invariant of the order $ k$
depends on the metric and the Riemann tensor $ R_{\na \nb \ng \d} $ and its
covariant derivatives of the order $\leq k$:
\BE
I=I(g_{\na \nb}, R_{\na \nb \ng \d}, \dots ,R_{\na \nb \ng \d; \ne_1 \dots \ne_k}).       
\EE
It has been known from 1902  what is the maximal number, $I(k,n)$,
of functionally independent invariants of the order $k$ in a Riemannian
space of dimension $n$. Denoting
\BE
D(k;n)=I(k;n)-I(k-1;n), 
\EE
Haskins \cite{Has}  found that
\BEA
D(0;n)={n \choose 3}\frac{n+3}{2} \quad \ \ {\rm for } \ k=0,  \\
 D(k;n)={n+k+1 \choose n+3}\frac{n(k+1)}{2} \quad \ \ {\rm for } \ k \geq 1. 
\EEA
Hence, in a 4-dimensional spacetime there exist
in general 14 functionally independent invariants of the order
zero which depend only on $g_{\na\nb}$ and $R_{\na\nb\ng\d} $, 60 invariants
depending on $g_{\na\nb}$, $R_{\na\nb\ng\d} $,
$R_{\na\nb\ng\d;\nr}$, 126 invariants
containing also $R_{\na\nb\ng\d;\nr\ns}$, etc. Clearly, when the derivatives
of the Riemann tensor are included, the number of independent invariants
grows rapidly. All 14 invariants of the zero order were explicitly given
in \cite{Geh}, their spinor equivalents can be found in \cite{WIT}.

In Section 2 we first briefly review basic definitions and relations
of the two-component spinor formalism and Newman-Penrose
formalism \cite{Newman}, which
will be needed later. Then we demonstrate the generally known fact that
in vacuum type {\it N} spacetimes with $\Lambda=0$ all zero-order invariants
vanish. If $\Lambda \not= 0$, some invariants are non-vanishing but they
are just constants depending on $\Lambda$. Using spinor formalism we prove
a helpful Lemma 1 on the properties of the invariants constructed from the
derivatives of the Weyl tensor.

In Section 3 we specialize to the type {\it N} vacuum solutions with $\Lambda$,
admitting a non-vanishing and non-twisting null geodesic congruence
(called Kundt's class in \cite{Kramer}). We prove that in these spacetimes all
invariants constructed from the Weyl tensor and its covariant
derivatives of arbitrary order vanish. The invariants constructed
from just the Riemann tensor are all constants depending on $\Lambda$.

Expanding and non-twisting type {\it N} solutions are discussed in
the first part of Section 4. Using again extensively the spinor
 and Newman-Penrose formalism we find that all the first-order
invariants vanish. However, the formalism indicates that a
non-vanishing second-order invariant may exist. A number of
attempts have eventually led to the non-vanishing invariant
\BE
I \ = \ R^{\na \nb \ng \d ;\ne\phi} R_{\na\nm\ng\nn;\ne\phi} R^{\nl \nm \nr \nn ; \ns \nt}
R_{\nl \nb \nr \d ; \ns \nt }. \label{invuv}
\EE
The zero and first-order invariants vanish also in the twisting case.
The invariant (\ref{invuv}) remains non-vanishing.

This invariant is then, in Section 5, used to analyze the nature
of approximate solutions \cite{Step2} , \cite{Finley}  we mentioned earlier. We find that
Stephani's conclusion, based on the linearized theory, remains true
for the third order solution obtained by Finley \etal: both
solutions contain singularities at large $r$. This raises more doubts
about the physical meaning of type {\it N} twisting solutions as describing
radiation fields outside bounded sources; nevertheless a definitive
statement can be made only when a general, exact solution is found.

The technique by which we arrive at the non-vanishing invariant
of the second order could probably be employed also in more complicated
cases of algebraically special spacetimes. Scalar invariants obtained
might play a role not only in classical relativity but also in a
quantum context.
\section{Higher-order curvature invariants: general properties}
\setcounter{equation}{0}
In this chapter we prove a lemma about the properties of the
curvature invariants of higher order in Petrov type {\it N} spacetimes.
The lemma will be used in the next section in which curvature invariants in the
specific classes of type {\it N} spacetimes will be analyzed. First, we have to
summarize some basic notations and relations which will be needed
later.
\hyphenation{formalism}
\subsection{Spinors and the Newman-Penrose formalism}
\hyphenation{for-mal-ism}
Spinors and the Newman-Penrose formalism have been reviewed by many
authors (see e.g.  \cite{Kramer},
\cite{Stew}, \cite{Brand}).
Here we give only a very brief
summary.

Consider a null congruence of geodesics with the tangent null vector
$l^{\na}$,
\BE l_{\na} l^{\na} = 0, \EE
which is affinely parametrized,
\BE l^{\na} l^{\nb} _{\ ;\na} = 0. \label{afin} \EE
(Hereafter, we assume Eq. (\ref{afin}) to be satisfied.)
The congruence is characterized by expansion $\theta$, twist
$\omega$ and shear $|\sigma|$ given by
\BEA
 \theta & = & \frac{1}{2} l^{\na}_{\ ;\na} \ , \label{exp}\\
 \omega & = & \sqrt{\frac{1}{2} l_{[\na;\nb]}l^{\na;\nb} }
      \ ,\label{twist} \\
|\ns| & = &  \sqrt{\frac{1}{2} l_{(\na;\nb)}
l^{\na;\nb} -\theta^2 }\ .\label{shear}
\EEA

In algebraically special spacetimes the shear vanishes.
A 2-component spinor field
$\sod{A}  \in W \  (A=1,2, \ W $ is a
2-dimensional complex vector space) can be associated with
 a null vector field $l^{\na}$ in a standard way,
\BE
  l^{\na} \vs \sou{A} \csou{A} ,  \label{korespl}
\EE
where $\csou{A} \in \bar W $, and another spinor field, $\siu{A} $, exists
 such that together with $\sou{A} $ it forms  a spinor basis satisfying
\BE
\sod{A} \siu{A} = \ospd{\eps}{AB} \sou{A} \siu{B} = 1, \ \
\sod{A} \sou{A} = \sid{A} \siu{A} = 0, \label{spprod}
\EE
where the Levi-Civita alternating symbol  $\ospd{\eps}{AB} $
 plays the role of the metric in spinor calculus.\

The  complex  null tetrad,
$\{ l^{\na} , n^{\na} ,m^{\na} ,{\bar{m}}^{\na} \}$,
in which $ l^{\na} $ is introduced by (\ref{korespl}) and
\BE
n^{\na} \vs \siu{A} \csiu{A}  , \
  m^{\na} \vs \sou{A} \csiu{A}  , \
  {\bar{m}}^{\na} \vs \siu{A} \csou{A} ,
\EE
satisfies the usual relations
\BE
 l_{\na} n^{\na} = -{\bar{m}}_{\na}  m^{\na} = 1. \label{nulbaz}
\EE
The (covariant) derivative operator $\nabla_{\na} $ can
be expressed in the form
\BE
\nabla_{\na} = n_{\na} D + l_{\na} \T - {\bar m}_{\na} \d -m_{\na} \cd ,
\label{rozlamb}
\EE
where
\BE
\fl
D = l^{\na} \nabla_{\na} \ ,\qquad  \T = n^{\na} \nabla_{\na} \ ,\qquad
\d = m^{\na} \nabla_{\na} \ ,\qquad \cd = {\bar m}^{\na} \nabla_{\na}
\ . \label{derdef}
\EE
In terms of the covariant derivative with spinor indices,
\BE
\dsd{A}{B} = \sgd{A\t{B} }{\na}  \nabla_{\na}
\EE
($\ns^{0} $ and $\ns^{i} $ are proportional to the unit and Pauli matrices),
  we have equivalently to Eqs. (\ref{derdef})
\BE
\eqalign{
D  = \sou{A} \csou{A} \dsd{A}{A}  &, \
\T = \siu{A} \csiu{A} \dsd{A}{A}  , \\
\d = \sou{A} \csiu{A} \dsd{A}{A}  &, \
\cd =\siu{A} \csou{A} \dsd{A}{A}  .\label{sdst} }
\EE
The Newman-Penrose twelve complex scalar spin coefficients
$\nk, \ne, \np \dots $ are defined as frame components of
the covariant derivatives of the null-tetrad vectors. Our notation follows
 that  customarily used in literature \cite{PR}, \cite{Stew}. Some
details are spelled out in Appendix A.

The Weyl tensor $C_{\na \nb \ng \d} $ in the spinor form is given by
\BE
 C_{\na \nb \ng \d} \ \vs \ \ospd{\Psi}{ABCD} \epsd{\t{A} \t{B} } \epsd{\t{C}
\t{D} } + \ospd{\bar \Psi}{\t{A} \t{B} \t{C} \t{D} }  \epsd{AB}
\epsd{CD} \ , \label{Wtenz}
\EE
where $ \ospd{\Psi}{ABCD} =\ospd{\Psi}{(ABCD)} $.
In the Newman-Penrose formalism the Weyl tensor is described by five
complex scalar quantities $ {\Psi}_{0},\ {\Psi}_{1},\
{\Psi}_{2}, \ {\Psi}_{3},\ {\Psi}_{4} $
given by the projections of $ {\Psi}_{ABCD} $
onto basis spinors $\sou{A} ,\siu{A} $.
The Riemann tensor $R_{\na \nb \ng \d} $ in the spinor form is given by
\BE
\eqalign{\fl
R_{\na \nb \ng \d} \ \ \vs \ospd{X}{ABCD} \epsd{\t{A} \t{B} } \epsd{\t{C}
\t{D} } + \ospd{\bar X}{\t{A} \t{B} \t{C} \t{D} }  \epsd{AB}
\epsd{CD}  \\ +
 \ospd{\Phi}{AB \t{C} \t{D} } \epsd{\t{A} \t{B} }
\epsd{CD} + \ospd{\bar \Phi}{\t{A} \t{B} CD} \epsd{AB}
\epsd{\t{C} \t{D} } , \label{rozpR}         }
\EE
where
\BEAH
\ospd{X}{ABCD}    =    \ospd{\Psi}{ABCD}    +   \frac{R}{12}
(\epsd{AC} \epsd{BD} +\epsd{AD} \epsd{BC} ),
\EEAH
$R$ is the scalar curvature, and the spinor
$ \ospd{\Phi}{AB \t{C} \t{D} } = \ospd{\Phi}{(AB)(\t{C} \t{D} ) }
= \ospd{\bar \Phi}{AB \t{C} \t{D} } $ corresponds  to the traceless
Ricci  tensor $ S_{\na \nb}  =  R_{\na \nb} - \frac{1}{4} R g_{\na \nb}
$:
\BE
\ospd{\Phi}{AB \t{A} \t{B} }   \vs  -\frac{1}{2}  S_{ab}.
\label{sbsRic}
\EE
In vacuum spacetimes with generally non-vanishing cosmological constant
$ \Lambda $,
\BE
R = 4 \Lambda,  \label{konstkriv}
\EE
and
\BE
\ospd{\Phi}{AB \t{C} \t{D} } = 0.
\label{nulPhi}
\EE
Bianchi identities connect the derivatives of $ \Psi $'s with the
$ \Psi $'s themselves and with the Newman-Penrose spin coefficients.
All Bianchi identities, the relations giving the
Riemann (Weyl) tensor in terms of the spin coefficients and the
commutation relations for the derivative operators (\ref{derdef})
are explicitly written down in  \cite{Kramer}, for example.

Here we are interested in the {\it vacuum} spacetimes of Petrov type
{\it N} with in general $\Lambda \not= 0 $. Let $\sou{A} $ be the 4-fold
principal null spinor of $\Psi_{ABCD} $, and $\siu{A} $ satisfies
(\ref{spprod}). Then we have
\BEA
\Psi_{0} =\Psi_{1} =\Psi_{2} =\Psi_{3} = 0, \label{nulkofPsi} \\
 \ospd{\Psi}{ABCD} = \Psi_{4} \sod{A} \sod{B} \sod{C} \sod{D} ,
\label{WspN}
 \EEA
and $R$ and $\ospd{\Phi}{AB \t{C} \t{D} } $ are given by Eqs.
(\ref{konstkriv}) and (\ref{nulPhi}).
The Bianchi identities reduce to
\BEA D\Psi_4 &=& (\nr-4\ne)\Psi_4 ,\label{Bid1} \\
    \d\Psi_4 &=& (\nt-4\nb)\Psi_4 ,\label{Bid2} \\
      \ns &=& \nk \ = \ 0 . \label{Bid3}
\EEA
Although under the conditions (\ref{konstkriv})--(\ref{WspN}) and
(\ref{Bid3}) the remaining Newman-Penrose equations take  much simpler
form than in general case, they still represent the set of 21 equations.
These are given in the Appendix.

As is well-known, transformations preserving the direction of $l^{\na} $
can be divided into two subclasses:
\begin{enumerate}
   \item
    Null rotations around vector $l^{\na} $
\BE
\fl
\eqalign{
{l'}^\na = l^\na \ , \quad {m'}^{\na} =m^\na + {\bar c}l^\na \ , \quad {n'}^\na =
n^\na + cm^\na +\bar c {\bar m}^{\na} +c \bar c l^\na ,\\
\ospu{o'}{A} = \sou{A} \ ,\qquad \ospu{\iota'}{A} = \siu{A}
+c\sou{A}  .\label{rotkoll} }
\EE
\item
Boosts in the $(l^{\na}, n^{\na})$ - plane and spacelike rotations in the
$ (m^{\na},{\bar m}^{\na}) $ - plane
\BE
\fl
\eqalign{
{l'}^{\na} = a^2 l^\na \ ,\quad {n'}^\na = a^{-2} n^{\na} \ , \quad {m'}^\na =
e^{2i \theta} m^\na , \\
 \ospu{o'}{A} = z \sou{A} \ , \quad
 \ospu{\iota'}{A} = {z}^{-1} \siu{A} \ ,
 \label{boostrot}  }
\EE
where $z = a e^{i \theta }$ .
\end{enumerate}
In the following we shall in particular need the behaviour of various
quantities under the pure constant boosts in the $(l^{\na}, n^{\na})$ -
plane ($a=$const, $\theta=0$ in (\ref{boostrot})). \\
If a quantity $\Omega$ transforms under the boosts as
\BE
\Omega'=a^q \Omega , \label{trOmg}
\EE
the number $q$ is called the boost-weight of $\Omega$.
We write $b(\Omega)$ to denote the boost-weight.
First notice that $b(\sou{A} )=1$, $b(\siu{A} )=-1$, and recall that
operators $D,\T, \d$ and $\cd$ are also weighted. The boost-weights
of the Newman-Penrose coefficients ($NP$) and
operators ($OP$)
are summarized in the following Table 1:
\begin{center}
\begin{tabular}{|c|c|c|}
\hline
$ x \in NP $ & $ x \in OP $ & $ b(x)=b(\bar{x}) $ \\
\hline
$ \nk $ & & 4 \\
\hline
$ \ne \ \ \ \nr \ \ \ \ns $ & $D$ & 2 \\
\hline
$ \na \ \ \ \nb \ \ \ \np \ \ \ \nt $ & $\d $ & 0\\
\hline
$\ng \ \ \ \nl \ \ \ \nm $ & $\T $ & -2 \\
\hline
$\nn$ & & -4 \\
\hline
\end{tabular}
\end{center}
We shall also need the behaviour of $\Psi_4 $:
\BE
{\Psi'}_4=a^{-4} \Psi_4  \Longrightarrow
   b({\Psi_4})= -4.    \label{bPsi4}
\EE

\subsection{Higher-order curvature invariants}

As noted in the Introduction, in a general spacetime there exist 14
independent invariants of the zero order, i.e.,  invariants depending only
on the metric and the Riemann tensor. In terms of the spinors
determining the
Riemann tensor according to  relations (\ref{Wtenz}), these invariants
can be constructed as products of the form
\BDM
\fl
\ospd{ \Phi}{\t {A} \t{B} MN}   \ospu{ \Phi}{\t {A} \t{B} MN} ,
\quad
\ospd{X}{ABMN} \ospu{X}{ABMN} , \quad
\ospu{\Phi}{GD \t{A} \t{B} } \ospd{\Phi}{RS \t{A} \t{B} }
\ospu{X}{RSKL} \ospd{\Phi}{KL \t{M} \t{N} } \ospdu{\Phi}{GD}{\ \
\ \t{M} \t{N} }  ,\nonumber
\EDM
etc. Since, however, in the vacuum Petrov type {\it N} spacetimes with $\Lambda$
equations (\ref{konstkriv})-(\ref{WspN}) are satisfied, it is easy to see
that among the zero-order invariants 9 vanish and 5 are dependent just
on the value of $\Lambda$ (as, e.g., (\ref{konstkriv})).
Hence, we have to turn to the invariants of higher order.

First, let us decompose the spinor derivative $\nderPsi$ into
the spinor basis of the appropriate spinor space.
Let us write
\BDM
{\cal W}^{[p,k]} \equiv
  \underbrace{W \times W \times \dots \times W}_{p}
 \times
 \underbrace{\Bar{W} \times \Bar{W} \times \dots \times \Bar{W}
}_{k} .
\EDM
Thus, ${\cal W}^{[p,k]}$ is $2^{p+k}$-- dimensional complex vector space whose
basis spinors $B_{1}^{[p,k]},
 B_{2}^{[p,k]}, \dots B_{2^{p+k}}^{[p,k]} $ can be constructed from
the tensorial products of $\sou{A} $ and $\siu{A} $ of the form
$\ospu{\xi_1}{A_1}  \ospu{\xi_2}{A_2}  \dots
   \ospu{\xi_{p}}{A_{p} }
  \ospu{\nl_1}{\dot X_1}  \ospu{\nl_2}{\dot X_2}  \dots
    \ospu{\nl_{k}}{\dot X_{k} }  $, where
 $\ospu{\xi_i}{A_i} $'s are $   \sou{A_i} $ or $\siu{A_i} $
  and
 $\ospu{\nl_j}{ \dot X_j } $'s are
$\csou{X_j} $  or  $ \csiu{X_j} $.
 Then the decomposition of  the $n$-th spinor derivative of
$\Psi_4 \sou{A} \sou{B} \sou{C} \sou{D} $ reads
\BE
\nderPsi = \sum_{i=1}^{2^{2n+4}} c_{i} B_{i}^{[n+4,n]} \ .
\label{rozklnderPsi}
\EE

We shall need to know some restrictions on the coefficients $c_i$
rather than their specific values. Let us first note that the
coefficients $c_i$  must all be the sums of the terms of the form
\BE
  X_1 X_2 \dots X_n \Psi_4 ,
 \label{tvarCi}
\EE
where $ X_i  \in NP $ or $ X_i  \in OP $. This can easily be seen by
regarding the well-known decompositions of $\dsu{A}{X} $
in terms of the NP~operators (\ref{derdef}) and of
 $\dsu{A}{X} \sou{B} $,
 $\dsu{A}{X} \siu{B} $ in terms of the basis spinors $o$, $ \iota$
and the NP~coefficients.

It is now useful to introduce a simple notation. If a product of
basis spinors has the form
$
Y \equiv \underbrace{\sou{A_1} \dots  \sou{A_{m_1} } }_{m_1}
         \underbrace{\csou{X_1} \dots  \csou{X_{m_2} } }_{m_2}
         \underbrace{\siu{B_1} \dots  \siu{B_{n_1} } }_{n_1}
         \underbrace{\csiu{Y_1} \dots  \csiu{Y_{n_2} } }_{n_2}
$
, then ${P_o}(Y) = m_1 + m_2 $ will denote the number of
$\sou{A} $'s and $\csou{A} $'s
, and ${P_{\iota}}(Y)=n_1+n_2$ of  $\siu{A} $'s
 and $\csiu{A} $'s which are contained in $Y$.

Now it is easy to see (essentially as a consequence of (\ref{spprod})),
that  if a spinor $ \ospS  \in {\cal W}^{[m,k]} $ has the form
\BE
\ospS = \sum_{i=1}^{2^{m+k}} s_i B_i^{[m,k]} \label{rozklS} ,
\EE
then all invariants formed from the products of $S^{\dots}$ must vanish
provided that the coefficients in (\ref{rozklS}) are such that
$s_i = 0$ for all $i$ for which
\BE
{P_o}(B_i^{[m,k]}) \leq {P_{\iota}}(B_i^{[m,k]}) . \label{ineqoi}
\EE
This observation enables us to prove the following \\[1mm]
{\bf Lemma 1:} \\[1mm]
Let an invariant constructed from the products of the  spinors \\
$\nderPsi$, for fixed $n$,  be non-vanishing. Then there exists a
non-vanishing quantity $X_1 X_2 \dots X_n \Psi_4 $,
$X_i \in NP \cup OP $,  such that
\BDM b(X_1 X_2 \dots X_n \Psi_4 ) =
\sum_{i=1}^{n}  b(X_i)  + b(\Psi_4) \geq 0 , {\mbox { i.e.}}, \
\sum_{i=1}^{n}  b(X_i)  \geq  4.
\EDM
{\it Proof:}
The $n$-th spinor derivative is of the form (\ref{rozklnderPsi}),
where the coefficients $c_i$  are sums of the terms of the form
(\ref{tvarCi}). According  to the observation above, an invariant
formed out of these derivatives will be non-vanishing only if
there exists  $c_i \not= 0$ such that ${P_o}(B_i^{[n+4,n]}) \leq
{P_{\iota}}(B_i^{[n+4,n]}) $. Since the basis spinors have the
boost-weight $d \equiv b(B_i^{[n+4,n]})=P{_o}(B_i^{[n+4,n]}) -
{P_{\iota}}(B_i^{[n+4,n]}) $, and since from (\ref{boostrot})
 and (\ref{bPsi4}) it follows that $b(c_i)=-d$, we see that $c_i$
must satisfy the condition $b(c_i) \geq 0$. Hence, there must exist
a non-vanishing quantity $X_1 X_2 \dots X_n \Psi_4 $,
$X_i \in NP \cup  OP $,   such that
$b(X_1 X_2 \dots X_n \Psi_4 ) \geq 0$.

\section{Non-expanding and non-twisting solutions}
\setcounter{equation}{0}
Type {\it N} vacuum solutions with $\Lambda$ admitting a
non-expanding and non-twisting null geodesic congruence
belong to Kundt's class (see \cite{Kramer}, ch. 27, for details).
Since $\theta=\omega=0$ (cf. Eqs. (\ref{exp}), (\ref{twist})), the
NP~coefficient
\BE
\nr = \theta + i \omega = 0.  \label{nulrho}
\EE
In this chapter we shall prove that in this class all the
curvature invariants of any order vanish (or are constants
determined  by $\Lambda$), generalizing  so  results
 of \cite{Jord} and \cite{Schmidt} where only the plane-wave
metrics are considered.

Choose the null tetrad parallelly propagated along the null
congruence determined by the multiple principal null direction
of the Weyl tensor and parametrized by an affine parameter. Thus,
only $\Psi_4  \not= 0$ (cf. equation (\ref{nulkofPsi})) and NP~coefficients
 $\ns = \nk = \np = \ne = 0 $. The NP~equations,
given in Appendix  for a general type {\it N} vacuum spacetime with
$\Lambda$, then simplify  considerably. From all the NP~equations
we shall only need those containing the operator $D$:
\BEA
D \nt &=& 0 \label{NProvKundt1} \nonumber ,\\
D \na  &=& 0 \nonumber ,\\
D\nb &=&0 \nonumber ,\\
D\ng &=& \nt\na+\nct\nb
- \frac{R}{24} \label{KundtNP} ,\\
D\nl &=& 0 ,\nonumber \\
D\nm &=& \frac{R}{12} \nonumber ,\\
D\nn &=& \nct\nm+\nt\nl ,\nonumber
\EEA
and the commutators
\BEA
(\T D - D\T) &=&  (\ng+\ncg)D-\nt\cd - \nct\d
 \label{KundtKom1} ,\\
(\d D - D\d) &=&  (\nca+\nb)D . \label{KundtKom2}
\EEA
We shall also need  the Bianchi identity (\ref{Bid1}) which
now simply reduces to
\BE
  D\Psi_4=0.         \label{KundtBid}
\EE

The simple form of the above equations  suggests the following
notation: let ${\cal F}_k$ be the set of functions $f$ such that
$f \in {\cal F}_k \ \Longleftrightarrow \ D^k f = 0.   $
From the NP~equations (\ref{KundtNP}) we easily find that
\BE
\eqalign{
 \na,\ \nb,\ \nt,\ \nl \ \in {\cal F}_1 \nonumber ,\\
\ng,\ \nm \ \in {\cal F}_2 , \quad    \label{DkNP}
\nn \in {\cal F}_3 .  }
\EE
Using this, and employing the equations (\ref{KundtKom1}) and
 (\ref{KundtKom2}) for the commutators, we can prove the following \\[1mm]
{\bf Lemma 2:} \\[1mm]
Let $f \in \ {\cal F}_k $. Then
$
   \mbox{(i)  } \ \  \d f  \in  {\cal F}_k, \ \   \cd f \ \in \ {\cal F}_k
,$

  $  \mbox{(ii) } \quad \T f \in {\cal F}_{k+1}. $\\
{\it Proof:}
(i) can  easily  be proven by induction. Applying
(\ref{KundtKom2}) on $f_{1}$, we immediately get $D\d f_{1}=0$
$\Rightarrow \d f_{1}  \in  {\cal F}_1$. Assuming then
$\d f_{k}  \in  {\cal F}_k$ and applying (\ref{KundtKom2}) on
$f_{k+1}$, we find $D^{k+1}\d f_{k+1}=0$
$\Rightarrow $ $\d f_{k+1}  \in  {\cal F}_{k+1}$ .\\
In order to prove (ii), we first show, by using Leibniz's
formula, that
$ D^{k+1} (f_2 {f}_k ) \ = \ 0 $ for all  $ k \geq 1$.
Then (ii) can be proven by induction similarly as in (i)
(the commutator (\ref{KundtKom1}) now being used instead of
(\ref{KundtKom2})).

It will now be useful to associate the number $p$ with any
NP~coefficient $X$ which will indicate the behaviour of $X$
under the action of the operator  $D$. Let
\BE
p (X)\ =\ k-1 \ ,\mbox{ if } \quad X \ \in {\cal F}_{k} \ \mbox{ but }
 X \ \not\in {\cal F}_{k-1} .\label{zobrp}
\EE
If $X \in OP$, i.e., $X$ is one of the NP~operators, we define
(being motivated by Lemma~2)
$p(\T)=1$, $p(\d)=0$, $p(D)=-1$.  \\
The values of $p$ for all relevant quantities are summarized in
the following Table 2:
\begin{center}
\begin{tabular}{|c|c|c|}
\hline
$ X \in NP $ & $ X \in OP $ & $ p (X) $ \\
\hline
$ \nn $ & & 2 \\
\hline
$ \ng \ \ \ \nm $ &$\T$  & 1\\
\hline
$ \na \ \ \ \nb \ \ \ \nt \ \ \ \nl $ & $ \d $ & 0\\
\hline
  &$ D $ & -1\\
\hline
\end{tabular} \\[2mm]
\end{center}
The indicators $p$ enable us to formulate easily
{\bf Lemma 3:} \\[1mm]
Consider (as in Lemma 1) a quantity
$ X_1 X_2 \dots X_n $ where $X_i \in NP \cup OP $.$ \\[1mm]
\mbox{If } \sum_{i=1}^{n} p (X_i) < 0 ,\mbox{ then }
  X_1 X_2 \dots X_n \Psi_4 =0.
$ \\[1mm]
{\it Proof:}
From the Bianchi identity (\ref{KundtBid}) we have
$D \Psi_4 = 0$. Regarding Table 2 we observe that the condition
$\sum_{i=1}^{n} p (X_i) < 0 $  requires that with  any
$X_i$ having  a positive $p(X_i)$, the operator
$D$ must appear  at least $p(X_i)$-times among $X_1 \dots X_n$ since only $p(D)$ is negative
(for example, with any $\nn$,  $D$ must appear at least two-times );
 an additional $D$ has then to enter $X_1 \dots X_n$ in order
that $\sum_{i=1}^{n} p (X_i) < 0 $. This results in
$ X_1  \dots X_n \Psi_4 =0 $.

Combining Lemma 1 and Lemma 3 we can now prove the following \\[1mm]
{\bf Proposition 1:} \\[1mm]
In type {\it N} vacuum spacetimes with $\Lambda$ admitting a non-expanding
and non-twisting null geodesic congruence all $n$-th order invariants
formed from the products of spinors $\nderPsi$, with $n$ arbitrary
but fixed, vanish.\\
{\it Proof:}
According to Lemmas 1 and 3, a non-vanishing invariant will
exist only if there are $X_i \in NP \cup OP$, $i=1 \dots n$,
such that
\BE \sum_{i=1}^{n} b (X_i) \geq 4  \ \ \mbox{and} \ \
\sum_{i=1}^{n} p (X_i) \geq 0 .  \label{eqProof}
\EE
The values of these sums depend on how many times the specific
NP~coefficient or NP~operator enter $X_1 \dots X_n$.
Let $m_1$ denote the number of the coefficients $\nn$ and $\ncn$,
${m_2}\ -$ of $\ng,\ncg,\nm$,~and~$\ncm$, ${m_3}\ -$ of $\nl$~and~$\ncl$,
${m_4}\ -$  of $ \na, \nca, \nb, \ncb, \nt$~and~$\nct$,
${k_1}\ -$  the number of operators $D$, ${k_2}\ -$  of $\d$~and~$\cd$,
and ${k_3}\ -$  of  $\T$, which enter $X_1 \dots X_n.$ Then, regarding
Tables 1 and 2, we find out that the inequalities (\ref{eqProof})
read
\BEAH
  &-4m_1 -2m_2 -2m_3 -2k_3 +2k_1 \geq 4 \mbox{, and } \
  & 2m_1 + m_2 +k_3 - k_1 \geq 0.
\EEAH
Combining the last two inequalities, we obtain $m_3 \leq -2$.
This is impossible since all $m$'s and $k$'s must be
non-negative.

From  Lemmas 1 and 3, and from the last proof it follows that
all the coefficients $c_i$ in the decomposition (\ref{rozklnderPsi})
which multiply basis spinors satisfying the condition (\ref{ineqoi}),
must necessarily vanish.
Hence, all non-vanishing terms in the decomposition
(\ref{rozklnderPsi}), with $n$ arbitrary, contain a larger number
of spinors $\sou{A} $  than of $\siu{A} $. Therefore, regarding
(\ref{spprod}) it is evident that even all invariants formed from
the products of the derivatives (\ref{rozklnderPsi}) with
{\it different} $n$'s must necessarily vanish. Recalling
the relations (\ref{Wtenz}) and (\ref{WspN}) between the Weyl
tensor $C_{\na \nb \ng \d}$, the Weyl spinor $\Psi_{ABCD}$
and the NP~scalar $\Psi_4$, we can now formulate our two
basic propositions.\\[1mm]
{\bf Proposition 2:} \\[1mm]
In type {\it N} vacuum spacetimes with $\Lambda$, admitting a
non-expanding and non-twisting null geodesic congruence,
all invariants constructed  from the Weyl tensor and its
covariant derivatives of arbitrary order vanish.\\[2mm]

Next, recall that in vacuum  spacetimes with $\Lambda$
Einstein's equations imply $R_{\na \nb} = \Lambda g_{\na \nb}$,
so that
\BE
C_{\na \nb \ng \d ; \ne} =  R_{\na \nb \ng \d ; \ne} .
\label{dCeqdR}
\EE
Then, regarding the spinor  form (\ref{rozpR}) of the
Riemann tensor we see that all invariants constructed from the Riemann
tensor and its covariant derivatives of arbitrary
order can be expressed in terms of the spinors
\BEA
&& \Psi_4 \sou{A} \sou{B} \sou{C} \sou{D} + \frac{\Lambda}{3}
(\epsu{AC} \epsu{BD} + \epsu{AD} \epsu{BC} ) ,\label{rRiem}\\
 && \dsu{C_1}{X_1} (\Psi_4 \sou{A} \sou{B} \sou{C} \sou{D} ) ,
\label{rWeyl}\\
&&\ \vdots \nonumber \\
&&\nderPsi .  \label{rndWeyl}
\EEA
Applying the considerations above and recalling that
$  \epsu{AB} = \sou{A} \siu{B}  - \sou{B} \siu{A} $,
we easily make sure that all terms in the invariants
containing  $\Psi_4 \sou{A} \sou{B} \sou{C} \sou{D} $ or
their derivatives of arbitrary order, vanish. The only
non-vanishing quantities can be formed from the constant
term $\frac{\Lambda}{3}
(\epsu{AC} \epsu{BD} + \epsu{AD} \epsu{BC} )  $
and are  dependent on $\Lambda$ only. We thus finally arrive
at \\[1mm]
{\bf Proposition 3:} \\[1mm]
In type {\it N} vacuum spacetimes with $\Lambda$, admitting a
non-expanding  and non-twisting null geodesic congruence,
all invariants constructed from the  products of the
Riemann tensor and its covariant derivatives of arbitrary
order vanish provided they contain a derivative of the Riemann
tensor. The invariants constructed from the Riemann tensor
itself are all constants depending on $\Lambda$.\\[2mm]
If $\Lambda=0$, the Riemann tensor is equal to the Weyl
tensor and all the invariants vanish by Proposition 2.

\section{Expanding and twisting solutions}
\setcounter{equation}{0}
\def \dcxi {d \bar \xi}
\def \cxi {\bar \xi}
\def \cA {\bar A}
This section  is divided into two parts.  In the first, we
analyze type {\it N}  spacetimes with $\Lambda$ with expanding but
non-twisting null congruences. We shall show that invariants of the
zero and first order vanish (or are constants determined by
$\Lambda$) as in the non-expanding and non-twisting case. However,
we will succeed in finding a non-vanishing invariant of the second order
depending on the expansion and on $\Psi_4$. In the second
part, we shall show that the invariants of the zero and first order again
vanish, and we shall demonstrate how the non-vanishing invariant
is modified when there is a non-zero twist.

Before we turn to the details we wish  to point out the
main reason why, with the expansion present, a non-vanishing
invariant may exist: it is due to the Newman-Penrose equation for the expansion,
$D \nr = \nr^2.$ Now  $D^n \nr \not= 0$ for any
$n$, and one even cannot formulate Lemma 3, for example.

\subsection{Non-twisting case}
All metrics of vacuum type {\it N} spacetimes with  $\Lambda$
were given by Garc\' \ii a-D\' \ii az and Pleba\' nski \cite{GP}. In their coordinates,
suitable for our purposes, the metric reads
\BE
\fl
ds^2 = 2v^{2} d\xi  d{\bar  \xi} + 2v{\bar A} d\xi du +2vAd{\bar \xi}  du + 2 \psi
dudv + 2(A {\bar A} + \psi B)du^{2} , \label{GDPmetric}
\EE
where
\BEA
      A &=& \epsilon \xi - vf \ , \nonumber \\
      B &=& -\epsilon + \frac {v}{2} (f,_{\xi} + \bar f,_{\bar \xi})+
         \frac {\Lambda}{6}v^{2}\psi  \ ,  \nonumber \\
      \psi &=& 1+\epsilon \xi \bar\xi \ , \\
      \epsilon &=& -1,\  0,\ +1 . \nonumber
\EEA
It is useful to choose the null tetrad corresponding to the forms
\BE
\eqalign{
\omega^{1} &= vd{\bar \xi}+{\bar A} du ,  \\
\omega^{2} &= vd\xi +A du ,    \\
\omega^{3} &= \psi du  ,      \\
\omega^{4} &= -dv-Bdu . }
\EE
Using this tetrad one obtains the NP~coefficients as follows:
\BEA
&&\ns=\nk=\ne=\np = \nl=\na=\nb=0 ,\ \  \nonumber \\
&&\ng = \frac{f,_{\xi}}{2\psi} +\frac{\Lambda v}{6}, \ \
\nt = -\frac{\epsilon \xi}{\psi v},  \ \
\nr = \frac{1}{v} ,  \label{NPkoefRTN} \\
&&\nn = -\frac{f,_{\xi \xi}}{2 \psi}, \ \
\nm =  -\frac{\Lambda v}{6}, \ \
\Psi_{4} = -\frac{f,_{\xi \xi \xi}}{2 \psi v} .\nonumber
\EEA

In the following calculations the specific forms of  most
of the NP~coefficients are not important - we need to know only
the value  of $\nr=1/v$, and the fact which of the coefficients
vanishes. Then we can write down all the NP~equations by specializing
the NP~equations in the Appendix  to the present case. The Bianchi
identities (\ref{Bid1}), (\ref{Bid2})   simplify now to the form
\BE
D \Psi_4 = \nr \Psi_4 ,  \ \ \ \d \Psi_4 = \nt \Psi_4 .
\label{BidGDP}
\EE
Using these and the NP~equations we can  write the first
spinor derivative of the Weyl spinor as
\BE
\eqalign{\fl
\dsu{E}{F} (\Psi_4 \sou{A} \sou{B} \sou{C} \sou{D} ) \ = \
&- \  (\T + 4\ng) \Psi_4 \SSu{0}{ABCDE} \csou{F}  \  \\ &+
   \cd \Psi_4   \SSu{0}{ABCDE} \csiu{F} \ + \
\Psi_4 (\nt\csou{F} - \nr \csiu{F} ) \SSu{1}{ABCDE} ,
\label{GDPr1der}       }
\EE
where we have introduced quantities $\SSu{j}{ABCDE} $ which denote
the  symmetrized products of the basis spinors $\sou{A} $ and
$\siu{A} $, and the subscript "$[j]$" gives the number of
$\siu{A} $'s, entering $S^{\dots} $. (For example, $
\SSu{1}{AB} =\sou{A} \siu{B} + \sou{B}  \siu{A} $.) As usually,
${\bar S}^{\dots}$ denotes the complex conjugate to $S^{\dots}$, with
$[j]$ being  the number of $\csiu{A} $'s. From equation
(\ref{GDPr1der}) it is then clear that all the first-order
invariants of the Weyl tensor vanish since there are not
enough $\iota$'s to be combined with $o$'s, as  was discussed
in detail in the previous section.

Now consider the second derivatives. Since calculations become
rather lengthy we turn to the computer algebra package
{\it Maple V }. Again using the
NP~equations and Bianchi identities, we arrive at the following
results:
\BE
\eqalign{\fl
 \dsu{G}{H} \dsu{E}{F} (\Psi_4 \sou{A} \sou{B} \sou{C} \sou{D} )
\  = \ {\cal A}  \SSu{0}{ABCDEG} \cSSu{0}{\dot F \dot H}
\  + \  {\cal B}  \SSu{1}{ABCDEG} \cSSu{0}{\dot F \dot H}
  \\
\  + \  {\cal C}  \SSu{0}{ABCDEG} \cSSu{1}{\dot F \dot H}
\  + \  {\cal D}  \SSu{2}{ABCDEG} \cSSu{0}{\dot F \dot H}
\  + \  {\cal E}  \SSu{1}{ABCDEG} \cSSu{1}{\dot F \dot H}
  \\
\  + \  {\cal F}  \SSu{2}{ABCDEG} \cSSu{1}{\dot F \dot H}
\  + \  {\cal G}  \SSu{1}{ABCDEG} \cSSu{2}{\dot F \dot H}
\  + \  {\cal H}  \SSu{2}{ABCDEG} \cSSu{2}{\dot F \dot H} ,
\label{rozkl2derRTN}    }
\EE
where functions ${\cal A,B,C}$ ... read
\BEA
{\cal A} \ &=& \ \Bigl\{ (\T + 9\ng + \ncg) \T \ + \ 4(5\ng + \ncg + \T )\ng
\ -\ 5\nt\nn - \ncn \cd \Bigr\} \Psi_4  , \nonumber  \\
{\cal B} \ &=& \ \d \Bigl\{ (\T-4\ng) \Psi_4 \Bigr\} ,\nonumber  \\
{\cal C} \ &=& \ -\cd \Bigl\{ (\T+4\ng) \Psi_4 \Bigr\} , \nonumber  \\
{\cal D} \ &=& \ 2\Psi_4 \nt^2   , \label{koefABC} \\
{\cal E} \ &=& \ (D \T + 4\nr \ng )\Psi_4  , \nonumber  \\
{\cal F} \ &=& \ -2 \Psi_4 \nr \nt , \nonumber  \\
{\cal G} \ &=& \ -2\nr \cd \Psi_4 , \nonumber  \\
{\cal H} \ &=& \ 2\Psi_4 {\nr}^2  . \nonumber
\EEA

The above expression (\ref{rozkl2derRTN}) indicates that  a
non-vanishing invariant may exist. Looking at the last term
in equation (\ref{rozkl2derRTN}), which is proportional to the function
${\cal H}$, we see that it contains, in contrast to all other terms,
 the same number of $o$'s and $\iota$'s. Consequently, a combination
of such terms may give a non-vanishing result. One can make sure,
however, that simple squares or even cubes of such terms do not
work. After a number of unsuccessful attempts to construct a
non-vanishing expression we arrived at the invariant
\BE
I \ = \ R^{\na \nb \ng \d ;\ne\phi} R_{\na\nm\ng\nn;\ne\phi} R^{\nl \nm \nr \nn ; \ns \nt}
R_{\nl \nb \nr \d ; \ns \nt } ,      \label{tvarinv}
\EE
or, regarding (\ref{dCeqdR}), equivalently
\BE
I \ = \ C^{\na \nb \ng \d ; \ne \phi} C_{\na \nm \ng \nn ; \ne \phi} C^{\nl \nm \nr \nn ; \ns \nt}
C_{\nl\nb\nr\d;\ns\nt}  .     \label{tvarinvC}
\EE
In terms of spinors employed to arrive at the invariant, we
obtain
\BE
I = 4 {\cal H}^2 {\bar {\cal H}}^2  \SSd{2}{BGDH}   \SSu{2}{BGDH}
 \cSSd{2}{\t{B} \t{G} \t{D} \t{H} }
 \cSSu{2}{\t{B} \t{G} \t{D} \t{H} } ,
\label{RTNkratsi}
\EE
where we have used the relation
$
\SSu{2}{ABCDEF} \sid{E} \sid{F} = \SSu{2}{ABCD}
\label{RTNzkratka}
$.
Since a straightforward calculation gives
$
\SSu{2}{BGDH}   \SSd{2}{BGDH}    = 6,
$
 we find
\BE
I = 144 {\cal H}^2 {\bar {\cal H}}^2  \label{IpomH} ,
\EE
or, regarding (\ref{koefABC}) and (\ref{NPkoefRTN}), we finally
obtain
\BE
I=9(2{\nr})^8{\Psi_4}^2 {\bar \Psi_4}^2 ,  \label{InvrhoRTN}
\EE
and
\BE
I=144 \frac{f_{\xi \xi \xi}^2 {\bar f}_{\cxi \cxi \cxi}^{2}
}{\psi^4 {v}^{12}} . \label{InvfRTN}
\EE

\subsection{Twisting case}
Twisting type {\it N} vacuum spacetimes with $\Lambda$ are not known,
except for the Hauser solution.  With non-vanishing twist, $\nr$
becomes complex, $\nr=\theta+i\omega$, and in contrast to
non-twisting spacetimes, the NP~coefficients $\na$, $\nb$, $\nl$
in general do not vanish. Hence, the NP~equations become much
more complicated. However, inspecting the relation generalizing
(\ref{GDPr1der}), we again find out that the zero
and first order invariants vanish.  The calculations of the second
derivatives (\ref{rozkl2derRTN}) are only feasible using
a computer algebra package (here again {\it Maple V }).  Functions
 ${\cal A,B,C} \dots$ in
(\ref{koefABC}) become much more lengthy, however, the most relevant
function, ${\cal H}$, remains the same. Therefore, using (\ref{IpomH})
 we find  that the invariant (\ref{tvarinv}), resp.
(\ref{tvarinvC}), becomes
\BE
I=9 (2 \nr)^4 (2 \ncr)^4 {\Psi_4}^2 {\bar \Psi_4}^2 .
\label{Invtwist}
\EE

Before moving on to possible applications of the invariants, let us
note that their form in the NP~formalism (not in terms of the Weyl
or Riemann tensor) could have been anticipated by considering
the behaviour of the NP~quantities under the transformations
(\ref{rotkoll}) and (\ref{boostrot}). Taking into account only
undifferentiated NP~quantities and assuming a type {\it N} spacetime
with $\nr \not=0$, one finds that the quantities which are invariant
under (\ref{rotkoll}) and (\ref{boostrot}) must be functionally dependent
on  $ \nr^2  \ncr^2 \Psi_4 {\bar \Psi_4} $. All such invariants
can thus be written in terms of the invariant (\ref{tvarinv}).
Of course, expression (\ref{tvarinv}) is invariant under {\it any}
tetrad and coordinate transformation.

\section{Applications}
\setcounter{equation}{0}
The invariant (\ref{tvarinv}) can be used to study the occurrence
of singularities in various expanding type {\it N} spacetimes.

\subsection{Expanding, non-twisting spacetimes}
For example, in the Garc\' \ii a D\' \ii az -- Pleba\'nski spacetimes
with the metric (\ref{GDPmetric}), we find that the invariant
(\ref{tvarinv}), which now becomes equal to the expression
(\ref{InvfRTN}), is diverging if for
$ f_{\xi \xi \xi}^2 {\bar f}_{\cxi \cxi \cxi}^{2} \not=0$
there is  (i) $v=0$, (ii) $\psi=1+\epsilon \xi {\bar \xi} = 0 $,
i.e., if $ \epsilon = -1 $ and $\xi {\bar \xi} =1 $. It also
diverges when  $v$ and $\psi$ are finite but
$ f_{\xi \xi \xi}^2 {\bar f}_{\cxi \cxi \cxi}^{2} $
diverges. If $f$ is a polynomial quadratic in $\xi$, the
spacetime can be shown  to be flat or de Sitter (if $\Lambda
\not=0$).
If, however,
\BDM
       f(\xi,u) =  c_{0}(u) + c_{1}(u)\xi + \dots +
        c_{n}(u)\xi^n ,
\EDM
where $n \geq 6 $, then  $I$ is singular at $\xi=\infty$,
 ${\bar \xi}  = \infty$ for each  $\epsilon= -1, 0, 1 $.

\subsection{Type {\it N} twisting spacetimes}

Recently two papers appeared discussing  the physical
meaning of the vacuum Einstein equations of Petrov
type {\it N} with expanding and twisting null congruence.
Stephani \cite{Step2} has argued, within the framework of the linearized
theory, that these solutions  contain singular lines ('pipes')
in space which  at any time extend arbitrarily far away from
a possible insular source. A few months ago, Finley, Pleba\'nski and
Przanowski \cite{Finley}, using an iterative approach, tried to resolve
"Stephani's paradox" by constructing the solution up to the third
order in their approximation scheme. They conclude that "up to the
third order, there do exist acceptable, regular solutions". What
does our invariant ({\ref{Invtwist}}) say about  the nature of these
approximate solutions?

The standard form of the type {\it N} twisting vacuum solutions
(see e.g. \cite{Kramer}) reads (using  here the usual signature +2)
\BE
\eqalign{
{\mbox{d}}s^2 &= 2 { \omega^{1} }{ \omega^{2} }- 2 { \omega^{3} }{ \omega^{4} } , \\
{ \omega^{1} } &= -\frac{{\mbox{d}} \zt}{P \ncr} ={ {\bar \omega}^{2} } , \label{twmetr}\\
{ \omega^{3} } &= {\mbox{d}}u + L{\mbox{d}}\zt + \cL {\mbox{d}} \czt , \\
{\omega^{4} } &= {\mbox{d}}r + W {\mbox{d}} \zt + \Bar{W}  {\mbox{d}}\czt ,
}
\EE
where the real function $P(\zt,\czt,u) $ and complex function $L(\zt,\czt,u)$
appearing in the metric satisfy  the relations
\BE
\eqalign{
\nr  = \frac{-1}{r+i  \Sigma} \ , \quad  2i \Sigma=  P^2 (
\cder L  - \der \cL )  ,  \\
W =   \frac{L_{,u}}{\nr}+i\der  \Sigma  \ , \quad  \der
\equiv  \der_{\zt}  - L \der_u \label{twrce1} . }
\EE
The function $\nr$ is indeed the NP~coefficient used before.
Introducing a real function $V$ by putting  $P=V_{,u}$,  the field
equations can be written in the form
\BE
\eqalign{
  (\der \der  \cder \cder  V)_{,u} =  P^{-1} (\der  \der
V)_{,u} (\cder \cder V)_{,u} , \\
{\mbox{Im}}( \der \der \cder \cder V)=0  \ , \quad
\der [P^{-1} (\cder \cder V)_{,u}] = 0  \label{twrce2}  . }
\EE
The coordinate system and field equations are invariant under
gauge transformations \cite{Kramer}, \cite{Step2}
\BE
\zt'=f(\zt), \quad u'=F(\zt,\czt,u),  \quad r'=rF^{-1}_{,u}. \label{twtransf}
\EE
The NP~component of the Weyl tensor is given by
\BE
\Psi_4  \  =  \  P^2  \nr  \partial_u  [P^{-1}  (\cder \cder
V)_{,u}] \label{compPsi4} .
\EE
Under the gauge transformation (\ref{twtransf}), the quantities appearing
above transform as follows:
\BE
\eqalign{
\nr'=F_{,u} \nr, \quad \Sigma=F_{,u} \Sigma' ,  \\
\der'={f'}^{-1} \der , \quad P'=F_{,u}^{-1} |f'| P ,  \\
L'={f'}^{-1} (Lf_{,u}-F_{,\zt}) , \label{quantransf} \\
\Psi_4'=f'^2 {({\bar f'})}^{-2}  F_{,u}^{-2} \Psi_4 . }
\EE

Using the expression  (\ref{twrce1}) and (\ref{compPsi4}) for
$\nr$ and $\Psi_4$, we can easily  calculate the invariant
(\ref{Invtwist}) to obtain
\BE
\fl
I=9 (2 \nr)^4 (2 \ncr)^4 {\Psi_4}^2 {\bar \Psi_4}^2 =
 2304 \frac{P^8}{(r^2 + \Sigma^2)^6}
 \Bigl( \der_u [ P^{-1} (\cder \cder V)_{,u} ] \Bigr)^2
 \Bigl( \der_u [ P^{-1} (\der \der V)_{,u} ] \Bigr)^2 .
 \label{Linv}
\EE
It can easily be checked, by using relations ({\ref{quantransf}}), that  ({\ref{Linv}})
does not change under the gauge transformation ({\ref{twtransf}}).
Of course, in its original forms ({\ref{tvarinv}}),
({\ref{tvarinvC}}), $I$ is invariant
under {\it any} coordinate transformation.

 Now Stephani \cite{Step2} found the general solution of the linearized field
equations in the form (using Stephani's notation)
\BE
\eqalign{
& P=1+\frac{\zt \czt}{2} ,   \\
& L = B(\zt, \czt) + \frac{C(u,\czt)}{(1+\zt \czt /2)^2} +
\frac{\czt^2 D(u,\zt)}{2(1+\zt \czt /2)^2}-
\frac{\czt D_{,\zt}}{(1+\zt \czt /2)} + D_{,\zt \zt} ,
\label{resLin} }
\EE
where functions $C(u,\czt)$ and $D(u,\zt)$ are arbitrary, $B(\zt,\czt)$
has to satisfy Im$(B_{,\czt})=0$. In order that spacetime is not flat,
i.e. $\Psi_4 \not= 0$, the condition
\BE
D_{,uu\zt \zt \zt} \not= 0 \label{nonflat}
\EE
has to be satisfied.\\
Using Stephani's solution (\ref{resLin}) in (\ref{Linv}), we find
the invariant to be
\BE
I^L = 2304 \frac{P^8}{r^{12}} D_{,\zt \zt \zt uu}^{2}
{\bar D}_{,\czt \czt \czt uu}^{2} .   \label{Linv2}
\EE
Defining ${\tilde C} (u,\czt) \equiv C(u,\czt) {\czt}^{-2} $ and
 ${\tilde D} (u,\zt) \equiv D(u,\zt) {\zt}^{-2} $,  Stephani calculates
the expression
\BE
{P^2} L_{,u \czt} = \frac{2 \czt {\tilde C}_{,u} (u,\czt)}{1+\zt \czt /2} + \czt^2 {\tilde C}_{,u \czt} -
\frac{2 \zt {\tilde D}_{,u} (u,\zt)}{1+\zt \czt /2} - \zt^2 {\tilde D}_{,u \zt}
,
\label{Step1}
\EE
which is invariant under the gauge transformation ({\ref{twtransf}}). Since functions  $\tilde C$ and $\tilde D$
are analytic in $\zt$ and $\czt$, respectively, they will have singularities  in the plane
$(\zt, \czt)$, i.e. on the sphere (recall the standard convention
$\zt=\sqrt{2} {\tan{\frac{\theta}{2}}} e^{i\varphi} $).  The expression (\ref{Step1})
will be regular only if  $\tilde C_{,u} \sim  \czt^{-1} $ and $\tilde D_{,u} \sim  \zt^{-1} $.
However, in this case the spacetime  is flat.

Clearly, function $D(u,\zt)$  has singularities in the plane $(\zt,\czt)$ and the invariant
({\ref{Linv2}})
will diverge unless $D_{,uu} \sim a \zt^2 + b\zt + c $. (Notice that if  $ D \sim \zt^3 $ than the
invariant diverges due to $P^8$.) However, this implies  flat spacetime again.

As mentioned above, Finley, Pleba\'nski and Przanowski \cite{Finley} constructed the twisting and diverging
type {\it N} solution up to the third order of  an iteration procedure. Their final expressions for
functions entering our invariant ({\ref{Linv}}) read
\BE
\eqalign{
\Sigma & \approx  \frac{1-\frac{1}{2} \zt \czt}{1+\frac{1}{2} \zt \czt} Im(f_2) -
2Re(a^{(1)}) Im \Bigl(\frac{df^{(2)}}{du} \Bigr) , \\
\Psi_{4} & \approx  \frac{\bar a^{(1)}}{r} \Bigl(\frac{1+\frac{1}{2} \zt \czt}{\czt}\Bigr)^2 \frac{d^3 \bar f^{2}}{du^3} ,
}
\EE
where $a^{(1)}$ is a complex constant, $f_2 (u) \equiv f^{(2)}(u) + f^{(3)}(u)$, $f^{(2)}(u)$,
$ f^{(3)}(u)$ being arbitrary functions of $u$. The curvature is non-vanishing so far as  $f^{(2)}_{,uuu}
\not= 0.$ Both quantities, $\Sigma/r$ and $K/r^2$ (for $K$ see equation (5.9) in \cite{Finley}), considered in
\cite{Finley}, which are invariant under gauge transformations ({\ref{twtransf}}), are indeed regular;
 the NP coefficient $\rho $ (cf.({\ref{twrce1}}))  entering
our curvature invariant ({\ref{Invtwist}}) is also regular. However, $(\Psi_4 {\bar \Psi}_4)^2$
is {\it not} regular on the
$(\zeta, {\bar \zeta} )$ -- sphere unless  $ {\bar a^{(1)}} f^{(2)}_{,uuu} = 0 $
which corresponds to flat
spacetime.  Indeed, $(\Psi_4 {\bar \Psi}_4)^2 \sim {(1+\frac{1}{2} \zt \czt)^8}/{(\zt \czt)^4} $
diverges
at $\zt \czt \rightarrow \infty $ (which in standard convention, with
$\zt=\sqrt{2} e^{i\varphi}  \tan{ \frac{\theta}{2}} $, corresponds to $\theta=\pi)$.

Therefore, we find  that  Stephani's conclusion based on the linearized theory remains true for the third
order solution analyzed by Finley, Pleba\'nski and Przanowski. Although this raises more doubts about the
interpretation of type {\it N} twisting solutions as representing radiation fields outside bounded sources,
solutions of full Einstein's equations may perhaps bring us surprises.

\ack
We thank  Ji\v r\' \ii \  Podolsk\'y, Alena Pravdov\' a and Hans Stephani for useful discussions.
We also acknowledge the supports  from the grants GACR--202/96/0206 of the
Czech Republic and GAUK --230/96 of the Charles University, and the hospitality
of the Albert Einstein Institute, Potsdam.

\appendix
\section{The Newman--Penrose equations in type {\it N} spacetimes}
The NP~coefficients are defined by the following table (see e.g. \cite{Stew}),
in which "$\nabla$" denotes respectively $D$, $\T$, $\d$, $\cd$; the first
line gives the definition in terms of the basis spinors, the second in terms
of the null tetrad.
\begin{center}
\begin{tabular}{|c|c|c|c|}
\hline & & & \\[-3mm]
$\nabla$ & $\sou{A} \nabla \sod{A} $ & $\sou{A} \nabla \sid{A} =
\siu{A} \nabla \sod{A} $  &  $\siu{A} \nabla \sid{A} $  \\[1mm]
\hline & & & \\[-3mm]
 & $m^{a} \nabla l_{a}$ & $ \frac{1}{2} (n^{a} \nabla l_{a} -
{\bar{m}}^{a} \nabla m_{a} ) $ & $ - {\bar m}^{a} \nabla n_{a} $
\\[1mm] \hline & & & \\[-3mm]
$D   $ & $ \nk $ & $ \ne $ & $ \np $ \\[1mm]
\hline & & & \\[-3mm]
$\T  $ & $ \nt $ & $ \ng $ & $ \nn $ \\[1mm]
\hline  & & &\\[-3mm]
$\d  $ & $ \ns $ & $ \nb $ & $ \nm $ \\[1mm]
\hline & & & \\[-3mm]
$\cd $ & $ \nr $ & $ \na $ & $ \nl $ \\[1mm]
\hline
\end{tabular}   \\[5mm]
\end{center}
In the vacuum type {\it N} spacetimes the null tetrad can be chosen so that
\BEAH
 \Psi_0 = \Psi_{1} = \Psi_{2} = \Psi_{3} = 0 ,\\
 \ospd{\Phi}{A \t{A} B \t{B} } = 0 ,\ \ns = \nk = 0,
\EEAH
the NP equations are
\BEA
D \nr  &=& \nr^{2}  + (\ne+\nce)\nr  , \nonumber \\
D \nt &=& (\nt + \ncp)\nr + (\ne - \nce)\nt  ,\nonumber  \\
D \na - \cd \ne &=& (\nr + \nce - 2 \ne)\na -\ncb \ne +(\ne + \nr)\np ,\nonumber  \\
D\nb-\d \ne &=& (\ncr-\nce)\nb - (\nca-\ncp)\ne ,\nonumber  \\
D\ng-\T\ne &=& (\nt+\ncp)\na+(\nct+\np)\nb-(\ne+\nce)\ng-(\ng+\ncg)\ne+\nt\np
- \frac{R}{24} \nonumber  ,\\
D\nl-\cd\np &=& \nr\nl+\np^{2}+(\na-\ncb)\np-(3\ne-\nce)\nl  ,\nonumber  \\
D\nm-\d\np &=& \ncr\nm+\np\ncp-(\ne+\nce)\nm-\np(\nca-\nb) +
\frac{R}{12} ,\nonumber  \\
D\nn-\T\np &=& (\np+\nct)\nm+(\ncp+\nt)\nl+(\ng-\ncg)\np-(3\ne+\nce)\nn ,\nonumber  \\
\T\nl-\cd\nn &=& -(\nm+\ncm)\nl-(3\ng-\ncg)\nl+(3\na+\ncb+\np-\nct)\nn-\Psi_4
, \label{NProv17}  \\
\d\nr &=& \nr(\nca+\nb)+(\nr-\ncr)\nt  ,\nonumber  \\
\d\na-\cd\nb &=&\nm\nr+\na\nca+\nb\ncb-2\na\nb+
\ng(\nr-\ncr)+\ne(\nm-\ncm)+\frac{R}{24}  ,\nonumber  \\
\d\nl-\cd\nm &=& (\nr-\ncr)\nn+(\nm-\ncm)\np+\nm(\na+\ncb)+\nl(\nca-3\nb)  ,\nonumber  \\
\d\nn-\T\nm &=& (\nm^2+\nl\ncl)+(\ng+\ncg)\nm-\ncn\np+(\nt-3\nb-\nca)\nn  ,\nonumber  \\
\d\ng-\T\nb &=& (\nt-\nca-\nb)\ng+\nm\nt-\ne\ncn-\nb(\ng-\ncg-\nm)+\na\ncl ,\nonumber  \\
\d\nt &=& \ncl\nr+(\nt+\nb-\nca)\nt  ,\nonumber  \\
\T\nr-\cd\nt &=&-\nr\ncm+(\ncb-\na-\nct)\nt+
       (\ng+\ncg)\nr-\frac{R}{12}  ,\nonumber \\
\T\na-\cd\ng &=& (\nr+\ne)\nn-(\nt+\nb)\nl+(\ncg-\ncm)\na+(\ncb-\nct)\ng ,\nonumber
\EEA
and commutators
\BEA
(\T D - D\T) &=&  (\ng+\ncg)D+(\ne+\nce)\T-(\nt+\ncp)\cd - (\nct+\np)\d,
\ \ \ \label{Kom1} \nonumber  \\
(\d D - D\d) &=&  (\nca+\nb-\ncp)D-(\ncr+\ne-\nce)\d, \ \ \ \\
(\d\T-\T\d) &=& -\ncn D+(\nt-\nca-\nb)\T+\ncl\cd+(\nm-\ng+\ncg)\d,
\ \ \ \nonumber  \\
(\cd\d-\d\cd) &=& (\ncm-\nm)D+(\ncr-\nr)\T-(\nca-\nb)\cd-(\ncb-\na)\d.
  \nonumber
\EEA

\end{document}

%% file: makra1.tex
%
\def \dcxi {d \bar \xi}
\def \cxi {\bar \xi}
\def \cA {\bar A}
\def \ffi {\varphi}
\def \vs {\longleftrightarrow}
\def \Bar#1 { \overline{#1} }
\def \der {\partial}
\def \cder {\bar \partial}
\def \zt {\zeta}
\def \czt {\bar \zeta}
\def \cL {\bar L}
\def \mbd {{\mbox{d}}}
\def \eps {\varepsilon}
\def \BE {\begin{equation}}
\def \EE {\end{equation}}
\def \BEAH {\begin{eqnarray*}}
\def \EEAH {\end{eqnarray*}}
\def \BEA {\begin{eqnarray}}
\def \EEA {\end{eqnarray}}
\def \BDM {\begin{displaymath}}
\def \EDM {\end{displaymath}}
\def \na {\alpha}
\def \nca {\bar \alpha}
\def \nb {\beta}
\def \ncb {\bar \beta}
\def \ng {\gamma}
\def \ncg {\bar \gamma}
\def \ne {\varepsilon}
\def \nce {\bar \varepsilon}
\def \nk {\kappa}
\def \nck {\bar \kappa}
\def \nl {\lambda}
\def \ncl {\bar \lambda}
\def \nm {\mu}
\def \ncm {\bar \mu}
\def \nn {\nu}
\def \ncn {\bar \nu}
\def \np {\pi}
\def \ncp {\bar \pi}
\def \nr {\rho}
\def \ncr {\bar \rho}
\def \ns {\sigma}
\def \ncs {\bar \sigma}
\def \nt {\tau}
\def \nct  {\bar \tau}
\def \cd {\bar \delta}
\def \d {\delta}
\def \T {\Delta}
\def \cm {\bar m}
\def \dsu#1#2 {\nabla^{ {\mbox{ {\tiny $\!\!\!\! #1 \!\dot #2 $}\rm}}} }
\def \dsd#1#2 {\nabla_{\mbox{{\tiny $\!\!\! #1 \!\dot#2 $}\rm}}}
\def \sou#1 {o^{{\mbox{{\tiny $ #1 $}\rm}}}}
\def \sod#1 {o_{{\mbox{{\tiny $ #1 $}\rm}}}}
\def \csou#1 {{\bar o}^{{\mbox{{\tiny $ \dot #1 $}\rm}}}}
\def \csod#1 {{\bar o}_{{\mbox{{\tiny $ \dot #1 $}\rm}}}}
\def \siu#1 {\iota^{{\mbox{{\tiny $ #1 $}\rm}}}}
\def \sid#1 {\iota_{{\mbox{{\tiny $ #1 $}\rm}}}}
\def \csiu#1 {{\bar \iota}^{{\mbox{{\tiny $ \dot #1 $}\rm}}}}
\def \csid#1 {{\bar \iota}_{{\mbox{{\tiny $ \dot #1 $}\rm}}}}
\def \om#1  { {\bf \omega^{ \!\!\! {\hat {\mbox{ {\tiny #1 } } } }}} \rm}
\def \com#1  { {\bf {\bar \omega}^{ \!\!\! {\hat {\mbox{ {\tiny #1 } } } }}} \rm}

\def \ospu#1#2 {#1^{ {\mbox{ {\tiny $\!\!\! #2 $} \rm}}} }
\def \ospd#1#2 {#1_{ {\mbox{ {\tiny $\!\!\! #2 $} \rm}}} }
\def \ospud#1#2#3 {#1^{ {\mbox{ {\tiny $\!\!\! #2 $} \rm}}}
                     _{ {\mbox{ {\tiny $\!\!\! \ \ #3 $} \rm}}}}
\def \ospdu#1#2#3 {#1_{ {\mbox{ {\tiny $\!\!\!  #2 $} \rm}}}
                     ^{ {\mbox{ {\tiny $\!\!\! \ \ #3 $} \rm}}}}
\def \SSu#1#2  {\mathop{S^{\scriptscriptstyle #2}}
\limits_{\scriptscriptstyle [#1] \hfill}}
\def \SSd#1#2  {\mathop{S_{\scriptscriptstyle #2}}
\limits_{\scriptscriptstyle [#1] \hfill}}
\def \cSSu#1#2  {\mathop{{\bar S}^{\scriptscriptstyle #2}}
\limits_{\scriptscriptstyle [#1] \hfill}}
\def \cSSd#1#2  {\mathop{{\bar S}_{\scriptscriptstyle #2}}
\limits_{\scriptscriptstyle [#1] \hfill}}
\def \Lab {L^{ {\mbox{ {\tiny $\!\!\! A $} \rm}}}
            _{ {\mbox{ {\tiny $     \ B $} \rm}}} }
\def \Su#1#2 {S^{ {\mbox{ {\tiny $\!\!\! #1 $} \rm}}}_{#2}}
\def \Sd#1#2 {S_{ {\mbox{ {\tiny $\!\!\! #1 $} \rm}}}^{#2}}
\def \dWsp {\dsu{E}{F} (\Psi_{4} \sou{A} \sou{B} \sou{C} \sou{D} ) }
\def \P#1 {\Psi_{#1}}
\def \epsu#1 {\eps^{ {\mbox{ {\tiny $\!\!\! #1 $}\rm}}} }
\def \epsd#1 {\eps_{ {\mbox{ {\tiny $\!\!\! #1 $}\rm}}} }
\def \t#1 { \dot #1 }
\def \skd#1#2  {\delta^{ {\mbox{ {\tiny $\!\!\! #1 $} \rm}}}
                      _{ {\mbox{ {\tiny $\!\!\! #2 $} \rm}}} }
\def \kd#1#2 {\delta^{#1}_{#2}}
\def \sgu#1#2  {\sigma^{ {\mbox{ {\tiny $\!\!\! #1 $} \rm}}}
                    _{ #2 }}
\def \csgu#1#2  {{\bar \sigma}^{ {\mbox{ {\tiny $\!\!\! #1 $} \rm}}}
                    _{ #2 }}

\def \sgd#1#2  {\sigma_{ {\mbox{ {\tiny $\!\!\! #1 $} \rm}}}
                    ^{  #2 }}
\def \otud#1#2#3 { #1 ^{#2}_{\ #3}}
\def \otdu#1#2#3 { #1 _{#2}^{\ #3}}
\def \ospsmuu#1#2#3 {#1^{{\cal #2 }
                     { {\mbox{ {\tiny $\!\!\! #3 $} \rm}}}}}
\def \ospsmdd#1#2#3 {#1_{{\cal #2 }
                     { {\mbox{ {\tiny $\!\!\! #3 $} \rm}}}}}
\def \ospsmud#1#2#3 {#1^{\cal #2 }
                       _{\mbox{ {\tiny $ \ #3 $} \rm}}}
\def \ospsmdu#1#2#3 {#1_{\cal #2 }
                       ^{\mbox{ {\tiny $ \ #3 $} \rm}}}
\def \ospsmddu#1#2#3#4 {#1_{{\cal #2 }{\mbox{ {\tiny $ \!\! #3 $} \rm}}}
                       ^{\mbox{ {\tiny $ \ \ \ \ #4 $} \rm}}}
\def \nderPsi {\dsu{C_{n}}{X_{n}} \dots \dsu{C_{1}}{X_{1}}
(\Psi_{4} \sou{A} \sou{B} \sou{C} \sou{D} )  }
\def \ospS {\ospu{S}{A_1 \dots A_m \dot X_1 \dots \dot X_k } }
\def \ospSvar {\ospu{S}{B_1 \dots B_m \dot Y_1 \dots \dot Y_k } }